\documentclass[english,conference,a4paper]{IEEEtran}

\IEEEoverridecommandlockouts
\usepackage{amsmath,amssymb,bbm,stmaryrd,soul,MnSymbol}
\usepackage{bm}
\usepackage{tikz}
\usepackage{stfloats}
\usetikzlibrary{shapes,decorations,matrix,positioning}
\usepackage{pgfplots}
\usepackage{babel,flushend}
\usepackage{scalefnt}

\renewcommand{\epsilon}{\varepsilon}
\newcommand{\dr}{d_c}
\newcommand{\dl}{d_v}

\global\long\def\dl{\texttt{l}}
\global\long\def\dr{\texttt{r}}
\global\long\def\emap{\epsilon^{\mathrm{MAP}}}
\global\long\def\xmap{x^{\mathrm{MAP}}}
\newcommand{\sddots}{\smash{\ddots}}
\newcommand{\svdots}{\smash{\vdots}}

\newcommand{\x}{\bm{x}}
\newcommand{\y}{\bm{y}}
\newcommand{\z}{\bm{z}}
\newcommand{\A}{\bm{A}}

\newcommand{\f}{\bm{f}}
\newcommand{\g}{\bm{g}}

\providecommand{\norm}[1]{\left\lVert#1\right\rVert}

\providecommand{\abs}[1]{\left\lvert#1\right\rvert}
\newtheorem{theorem}{Theorem}
\newtheorem{lemma}{Lemma}
\newtheorem{definition}{Definition}
\newtheorem{example}{Example}
\newtheorem{remark}{Remark}
\newtheorem{corollary}{Corollary}

\pgfplotsset{tick label style={
font=\Large}}

\begin{document}
\pgfdeclarelayer{background}
\pgfdeclarelayer{foreground}
\pgfsetlayers{background,main,foreground}

\title{A Simple Proof of Threshold Saturation \\ for Coupled Scalar Recursions}

\author{Arvind Yedla, Yung-Yih Jian, Phong S. Nguyen, and Henry D. Pfister%
\thanks{This material is based upon work supported by the
  National Science Foundation (NSF) under Grants No. 0747470 and No. 0802124.
  Any opinions, findings, conclusions, and recommendations expressed
  in this material are those of the authors and do not necessarily
  reflect the views of these sponsors.}
\\Department of Electrical and Computer Engineering, Texas A\&M University
\\Email: \{yarvind,yungyih.jian,pnguyen,hpfister\}@tamu.edu}

\maketitle

\begin{abstract}
  Low-density parity-check (LDPC) convolutional codes (or
  spatially-coupled codes) have been shown to approach capacity
  on the binary erasure channel (BEC) and 
  binary-input memoryless symmetric channels.  The mechanism behind
  this spectacular performance is the threshold saturation phenomenon,
  which is characterized by the belief-propagation threshold of the
  spatially-coupled ensemble increasing to an intrinsic noise threshold
  defined by the uncoupled system.


In this paper, we present a simple proof of threshold saturation that applies to a broad class of coupled scalar recursions.
The conditions of the theorem are verified for the density-evolution (DE) equations of irregular LDPC codes on the BEC, a class of generalized LDPC codes, and the joint iterative decoding of LDPC codes on intersymbol-interference channels with erasure noise.
Our approach is based on potential functions and was motivated mainly
by the ideas of Takeuchi et al. 
The resulting proof is surprisingly simple when compared to previous methods.

\end{abstract}

\vspace{-0.25mm}
\section{Introduction}

Convolutional low-density parity-check (LDPC) codes, or spatially-coupled (SC) LDPC codes, were introduced in \cite{Felstrom-it99} and shown to have excellent belief-propagation (BP) thresholds in \cite{Sridharan-aller04,Lentmaier-isit05,Lentmaier-it10}.
Moreover, they have recently been observed to \emph{universally} approach the capacity of various channels~\cite{Lentmaier-it10,Kudekar-istc10,Rathi-isit11,Yedla-isit11,Kudekar-isit11-DEC,Nguyen-arxiv11,Nguyen-icc12,Kudekar-arxiv12}.

The fundamental mechanism behind this is explained well in \cite{Kudekar-it11}, where it is proven analytically for the BEC that the BP threshold of a particular SC ensemble converges to the maximum-a-posteriori (MAP) threshold of the underlying ensemble.
This phenomenon is now called \emph{threshold saturation}.
A similar result was also observed independently in \cite{Lentmaier-isit10} and stated as a conjecture.
The same result for general binary memoryless symmetric (BMS) channels was first empirically observed~\cite{Lentmaier-it10,Kudekar-istc10} and recently proven analytically~\cite{Kudekar-arxiv12}.

The underlying principle behind threshold saturation appears to be very general and it has now been applied, with much success, to a variety of more general scenarios in information theory and coding.
In~\cite{Hassani-itw10}, the benefits of spatial coupling are described for $K$-satisfiability, graph coloring, and the Curie-Weiss model in statistical physics.
SC codes are shown to achieve the entire rate-equivocation region for the BEC wiretap channel in~\cite{Rathi-isit11}.
The authors observe in~\cite{Yedla-isit11} that the phenomenon of threshold saturation extends to multi-terminal problems (e.g., a noisy Slepian-Wolf problem) and can provide universality over unknown channel parameters.
Threshold saturation has also been observed for the binary-adder channel~\cite{Kudekar-isit11-MAC}, for intersymbol-interference channels~\cite{Kudekar-isit11-DEC,Nguyen-arxiv11,Nguyen-icc12}, for message-passing decoding of code-division multiple access (CDMA)~\cite{Takeuchi-isit11,Schlegel-isit11}, and for iterative hard-decision decoding of SC generalized LDPC codes~\cite{Jian-isit12}.
For compressive sensing, SC measurement matrices were investigated first with verification-based reconstruction in~\cite{Kudekar-aller10}, and then proved to achieve the information-theoretic limit in~\cite{Donoho-arxiv11}.

In many of these papers it is conjectured, either implicitly or explicitly, that threshold saturation occurs for the studied problem.
A general proof of threshold saturation (especially one where only a few details must be verified for each system) would allow one to settle all of these conjectures simultaneously.
In this paper, we provide such a proof for systems with scalar density-evolution (DE) equations.

Our method is based on potential functions and was motivated mainly by the approach taken in~\cite{Takeuchi-arxiv11}.
It turns out that their approach is missing a few important elements and does not, as far as we know, lead to a general proof of threshold saturation.
Still, it introduces the idea of a potential function defined by an integral of the DE recursion and this is an important element in our approach.
More recently, a continuum approach to DE is used, in~\cite{Donoho-arxiv11}, to prove threshold saturation for compressed sensing and was reported informally to give a general proof~\cite{Richardson-pc11}.

\vspace{-0.25mm}

\section{A Simple Proof of Threshold Saturation}

In this section, we provide a simple proof of threshold
saturation via spatial-coupling for a broad class of scalar
recursions.  The main tool is a potential theory for scalar recursions
that extends naturally to coupled systems of recursions.

\subsection{Single-System Potential}
\label{sec:single-sytem-potent}

First, we define potential functions for a class of scalar recursions
and discuss threshold parameters associated with the potential.

\begin{definition}
  An \emph{scalar admissible system} $(f,g)$ parameterized by $\epsilon \in [0,1]$, is defined by the recursion
  \begin{align}
    \label{eq:1}
    x^{(\ell+1)} = f(g(x^{(\ell)});\epsilon),
  \end{align}
  where $f:[0,1]\times[0,1]\to[0,1]$ is strictly increasing in both
  arguments for $x,\epsilon\in(0,1]$, and $g:[0,1]\to[0,1]$ satisfies
  $g'(x) > 0$ for $x\in (0,1)$.  We also assume that $f(0;\epsilon) =
  f(x;0) = g(0) = 0$ and that $f,g$ have continuous second derivatives
  on $[0,1]$ w.r.t. all arguments.
\end{definition}
\begin{definition}
  The \emph{potential function} $U(x;\epsilon)$ of a scalar admissible system~$(f,g)$ is defined by
  \begin{align}
    \label{eq:2}
    U(x;\epsilon) &\triangleq \int_0^x\left(z - f(g(z);\epsilon)\right)g'(z)\mathrm{d}z \notag\\
    &= xg(x) - G(x) - F(g(x);\epsilon),
  \end{align}
  where $F(x;\epsilon) = \int_0^x f(z;\epsilon)\mathrm{d}z$ and $G(x) = \int_0^xg(z)\mathrm{d}z$.
\end{definition}

\begin{figure}
  \centering
  \input{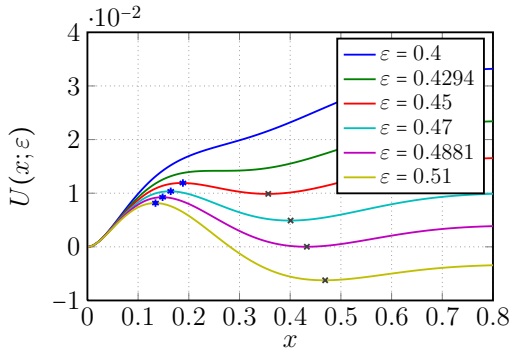}
\vspace{-1mm}
  \caption{The potential function of the (3,6)-regular LDPC ensemble is shown for a range of $\epsilon$.
    Here $\epsilon_s^* \approx 0.4294$, $\epsilon^* \approx 0.4881$, and the stationary points are marked.
    Notice that, for $\epsilon < \epsilon_s^*$, $U(x;\epsilon)$ has no stationary points.}
  \label{fig:single_potential}
\end{figure}

\begin{definition}
\label{def:fp_sp_eroot}
For $x,\epsilon\in[0,1]$, we have the following terms.
\begin{itemize}
\item
  For fixed $\epsilon$, $x$ is a \emph{fixed point} (f.p.) iff $x = f(g(x);\epsilon)$.
\item
  For fixed $\epsilon$, $x$ is a \emph{stationary point} (s.p.) if $U'(x;\epsilon) = 0$.
\item
  For $0<x\leq f(g(x);1)$, we define $\epsilon(x)$ to be the unique
  \emph{$\epsilon$-root} of the equation $x-f(g(x);\epsilon)=0$.
\end{itemize}
\end{definition}
\begin{lemma}
  The potential function of a scalar admissible system has the following properties:
  \begin{enumerate}
  \item $U(x;\epsilon)$ is strictly decreasing in $\epsilon$, for $\epsilon \in (0,1]$.
  \item An $x\in [0,1]$ is a f.p. iff it is a s.p. of the potential.
  \end{enumerate}
\end{lemma}
\begin{IEEEproof}
  These properties hold because the potential function is the
  integral of $\left(z - f(g(z);\epsilon)\right)g'(z)$ w.r.t. $z$, which is
  strictly decreasing in $\epsilon$, for $\epsilon \in (0,1]$, and
  zero iff $z$ is a fixed point of the recursion.
\end{IEEEproof}
\begin{definition}
  The \emph{single-system threshold} is defined to be
\begin{align*}
  \epsilon_s^* &= \sup \left\{ \epsilon\in [0,1] \mid U'(x;\epsilon) > 0 \; \forall \; x \in(0,1] \right\},
\end{align*}
and is the $\epsilon$-threshold for convergence of the single-system recursion to 0.
It is well defined because $U'(x;\epsilon)$ is strictly decreasing in $\epsilon$.
This implies that, for $\epsilon < \epsilon_s^*$, (\ref{eq:1}) has no fixed points in $(0,1]$.
For DE recursions associated with BP decoding, the threshold $\epsilon_s^*$ is called the BP threshold.
\end{definition}

\begin{example}
  For the standard irregular ensemble of LDPC codes (e.g., see~\cite{RU-2008}), the DE recursion,
  \[ x^{(\ell+1)} = \epsilon \lambda (1-\rho(1-x^{(\ell)})), \] is an
  scalar admissible system with $f(x;\epsilon) = \epsilon \lambda(x)$
  and $g(x)=1-\rho(1-x)$.  In this case, the single-system potential
  is given by~(\ref{eq:ldpc_potential}) and is shown in
  Fig.~\ref{fig:single_potential} for the $(3,6)$-regular LDPC code
  ensemble defined by $(\lambda,\rho)=(x^2,x^5)$.
\end{example}
\begin{definition}
\label{def: sxux}
For $\epsilon > \epsilon_s^*$, we define the \emph{minimum unstable} fixed point to be~\vspace*{-1mm}
\begin{equation*}
  u(\epsilon) = \sup\{\tilde{x}\in[0,1] \mid f(g(x);\epsilon)<x,x\in(0,\tilde{x})\}.
 \end{equation*}
\end{definition}
\begin{definition}
\label{def: PotentialThreshold}
Let the \emph{potential threshold} of the system be \vspace{-0.5mm}
\begin{equation}
  \label{eq:3}
  \epsilon^* = \sup \{\epsilon\in[0,1] \mid u(\epsilon)>0, \min_{x\in[u(\epsilon),1]} U(x;\epsilon) \geq 0\} \vspace*{-1mm}
\end{equation}
and $\Delta E(\epsilon) = \min_{x\in [u(\epsilon),1]} U(x;\epsilon)$
be the \emph{energy gap} of the system for $\epsilon \in (\epsilon_s^*,1]$.%
\end{definition}
\begin{remark}
  One consequence of this definition is that, if $\epsilon \! < \!
  \epsilon^*$, then $U(x;\epsilon) \! > \! 0$ for $x\in (0,1]$.
  Likewise, if $\Delta E(\epsilon) = 0$ and $u(\epsilon)\! >\! 0$, then $\epsilon \! = \!
  \epsilon^*$. For DE recursions associated with BP decoding, the
  potential threshold is analogous to the threshold predicted by the
  Maxwell conjecture~\cite[Conj.~1]{Measson-it08}.
\end{remark}

\subsection{Coupled System Potential}
\label{sec:coupl-syst-potent}

Now, we extend our definition of potential functions to coupled
systems of scalar recursions.  In particular, we consider a
``spatial-coupling'' of the single-system recursion, (\ref{eq:1}),
that gives rise to the vector recursion (\ref{eq:4}).  For the
vector recursion of the coupled system, we define a potential
function and show that, for $\epsilon < \epsilon^*$, the only fixed
point of the coupled system is the zero vector.

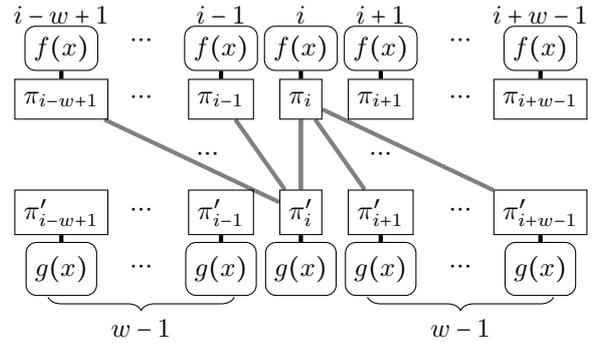
\begin{figure}
  \centering
  \begin{tikzpicture}[xscale=0.35,yscale=0.75]
  \foreach \i/\text in {2/i-w+1,8/i-1,11/i,14/i+1,20/i+w-1} { 
    \node[draw=black,rectangle,rounded corners]
    (f_node\i) at (\i,3.9) {$f(x)$}; 
    \node[minimum size=20pt,draw=black,rectangle,rounded corners] (g_node\i) at (\i,0)
    {$g(x)$};
    \node[draw,rectangle,minimum size=15pt] (p1_node\i) at (\i,3) {$\pi_{\text}$};
    \node[draw,rectangle,minimum size=15pt] (p2_node\i) at (\i,1)
    {$\pi_{\text}'$};
    \draw[ultra thick] (f_node\i) -- (p1_node\i);
    \draw[ultra thick] (g_node\i) -- (p2_node\i);
} 
\draw[decorate,decoration={brace,amplitude=5pt}]
  (8.5,-0.5) -- node[midway,below=5pt] {$w-1$} (1.5,-0.5);
\draw[decorate,decoration={brace,amplitude=5pt}] (20.5,-0.5) --
  node[midway,below=5pt] {$w-1$} (13.5,-0.5); 

\foreach \i/\text in {2/i-w+1,8/i-1,11/i,14/i+1,20/i+w-1} 
 \node at (\i,4.5) {$\text$};
 \begin{pgfonlayer}{background}
\foreach \i/\j in {11/2,11/8,11/11} {
\draw[ultra thick,gray] (p2_node\i) -- (p1_node\j);
}
\foreach \i/\j in {11/11,11/14,11/20} {
\draw[ultra thick,gray] (p1_node\i) -- (p2_node\j);
}
\node (tmp) at (7.5,2) {$\cdots$};
\node (tmp) at (14,2) {$\cdots$};
 \end{pgfonlayer}
\foreach \x in {5,17} {
\foreach \y in {4,3,1,0} { 
\node (blu) at (\x,\y) {$\cdots$};
}}
\end{tikzpicture}
\vspace{-1mm}
  \caption{A portion of a generic SC system.
The $f$-node at position~$i$ is coupled with the $g$-nodes at positions $i,\ldots,i+w-1$ and, by reciprocity,
$g$-node at position $i$ is coupled with the $f$-nodes at positions
$i-w+1,\ldots,i$. Here, $\pi_i$ and $\pi_i'$ are random permutations.}
  \label{fig:sc_system}
\end{figure}

\begin{definition}[cf.~\cite{Kudekar-it11}]
\label{def:SpatiallyCoupled}
  The basic \emph{spatially-coupled system} is defined by placing $2L+1$
  single systems at positions in the set
  $\mathcal{L}_0 \triangleq \{-L,-L+1,\ldots,L\}$ and coupling them with $w$ systems as shown in
  Fig.~\ref{fig:sc_system}.
  Let $x_i^{(\ell)}$ be the input to the $g$-function in the $i$-th position after $\ell$ iterations and define $x^{(\ell)}_i = 0$ for $i\notin\mathcal{L}\triangleq\{-L,-L+1,\ldots,L+w-1\}$ and all $\ell$.
  For the coupled system, we have the recursion \vspace{-1mm}
\begin{align}
\label{eq:4}
  x^{(\ell+1)}_i &=
  \frac{1}{w}\sum_{k=0}^{w-1}f\left(\frac{1}{w}\sum_{j=0}^{w-1}g(x^{(\ell)}_{i+j-k});\epsilon_{i-k} \right),
\vspace{-4mm}
\end{align}
where $\epsilon_{i} = \epsilon$ for $i\in \mathcal{L}_0$ and $\epsilon_i = 0$ for $i\notin \mathcal{L}_0$.
\end{definition}

\begin{definition}
  The recursion defined by~(\ref{eq:4}) can be rewritten as a
  \emph{vector recursion}.  Let $\f(\x;\epsilon)$ and $\g(\x)$ be
  defined for vector arguments by $[\f(\x;\epsilon)]_i =
  f(x_i;\epsilon)$ and $[\g(\x)]_i = g(x_i)$, respectively.
  Then,~(\ref{eq:4}) is equivalent to
\[  \x^{(\ell+1)} = \A_2 ^{\intercal}\f(\A_2  \g(\x^{(\ell)});\epsilon), \]
where $\A_2$ is the $(2L+1)\times(2L+w)$ matrix given by

\begin{center}
\hspace{9mm}
\begin{tikzpicture}[decoration=brace]
    \matrix (m) [matrix of math nodes,left delimiter=[,right delimiter={]}] {
        1 & 1 & \cdots & 1 & 0 & 0 & \cdots & 0\\
        0 & 1 & 1 & \cdots & 1 & 0 & \smash{\cdots} & 0 \\
        \vphantom{0}\smash{\vdots} & \smash{\ddots} & \smash{\ddots} & \smash{\ddots} & \smash{\ddots} & \smash{\ddots} & \smash{\ddots} & \smash{\vdots} \\
        0 & \cdots & 0 & 1 & 1 & \cdots & 1 & 0\\
        0 & \cdots & 0 & 0 & 1 & 1 & \cdots & 1\\
    };
    \node[transform canvas={xshift=-2.9cm},thick] (m-2-1.west) {$\A_2=\dfrac{1}{w}$};
    \draw[decorate,transform canvas={xshift=0.4cm},thick] (m-1-8.north east) -- node[right=2pt] {\rotatebox{270}{\small$2L+1$}} (m-5-8.south east);
    \draw[decorate,transform canvas={yshift=0.0em},thick] (m-1-1.north west) -- node[above=2pt] {\small$w$} (m-1-4.north east);
    \draw[decorate,transform canvas={yshift=-0.0em},thick] (m-5-8.south east) -- node[below=2pt] {\small$2L+w$} (m-5-1.south west);
\end{tikzpicture}
\vspace{2mm}
\end{center}
\end{definition}

\begin{remark}
In contrast to~\cite{Kudekar-it11}, the SC recursion defines $\x^{(\ell)}$ to be the \emph{SC-average} of the $f$-function (e.g., bit node in the LDPC example) output values rather than the output values.
Since there are $2L+1$ active $f$-function outputs, the vector $\x^{(\ell)}$ contains the $2L+w$ active averaged values after convolution with the length-$w$ averaging sequence.
This also shifts the maximum value of the vector from position 0, in~\cite{Kudekar-it11}, to position $i_0 \triangleq \left\lfloor\frac{w-1}{2}\right\rfloor$ in this work.

One might also expect that $A_2$ is square and that all rows and columns should sum to 1, but the termination leads to a rectangular matrix with reduced column sums near the boundaries.
In particular, $\A_2 \g (\x)$ is a length $2L+1$ vector representing the inputs to the $2L+1$ active $f$-functions.
\end{remark}

\begin{definition}[cf.~\cite{Kudekar-it11}]
\label{def:OneSidedSC}
  The \emph{one-sided spatially-coupled system} is a modification of (\ref{eq:4}) defined by fixing the values of positions outside $\mathcal{L}'\triangleq\{-L,L+1,\ldots,i_0\}$, 
  where $i_0 = \left\lfloor\frac{w-1}{2}\right\rfloor$ is the position of the maximum element of $\x^{(\ell)}$.
  It fixes the left boundary to zero by defining $x^{(\ell)}_i = 0$ for $i< -L$ and all $\ell$.
  It forces the right boundary to a floating constant by setting $x^{(\ell)}_i = x^{(\ell)}_{i_0}$ for $i\geq i_0$ and all $\ell$.
\end{definition}

\begin{definition}
Let the \emph{vector one-sided SC recursion} be \vspace{-1mm}
\begin{align}
  \label{eq:spatial_fp}
  \x^{(\ell+1)} = \A ^{\intercal}\f(\A  \g(\x^{(\ell)});\epsilon), \vspace{-1mm}
\end{align}
where $\x^{(\ell)}\!=\!
[x_{-L-w}^{(\ell)},\ldots,x_{2w+i_0}^{(\ell)}]$
and $\A$ is the $(L\!+\!3w\!+\!i_0\!+\!1)\times
(L\!+\!3w\!+\!i_0\!+\!1)$ matrix given by
\begin{align*}
  \A = \frac{1}{w}
   \begin{bmatrix}
    1      & 1      & \cdots & 1      & 0      & \cdots  & 0      \\
    0      & 1      & 1      & \cdots & 1      & \sddots & \svdots \\
    \svdots & \sddots & \sddots & \sddots & \sddots & \sddots & 0      \\
    0      & \cdots & 0      & 1      & 1      & \cdots & 1      \\
    0      & 0      & \cdots & 0      & 1      & \sddots & 1      \\
    0      & 0      & \cdots & 0      & 0      & 1      & \svdots \\
    0      & 0      & \cdots & 0      & 0      & 0      & 1      \\
   \end{bmatrix}.
\end{align*}
\end{definition}
\begin{remark}
  The right hand side of \eqref{eq:spatial_fp} accurately represents a single iteration of the one-sided SC system update for $i\in\mathcal{L}'$, but cannot be used recursively unless the boundary condition $x_i^{(\ell)} = x_{i_0}^{(\ell)}$ for $i\geq i_0$ is enforced after each step.
\end{remark}

\begin{lemma}[cf.~{\cite[Lem. 14]{Kudekar-it11}}]
\label{lem:monotone}
For both the basic and one-sided SC systems, the recursions are component-wise decreasing with iteration and converge to well-defined fixed points.
The one-sided recursion is also a component-wise upper bound on the basic SC recursion for $i\in\mathcal{L}$ and it converges to a non-decreasing fixed-point vector.

\end{lemma}
\begin{IEEEproof}
The proof follows easily from the arguments in \cite[Sec.~V]{Kudekar-it11} and is, hence, omitted.
\end{IEEEproof}

\begin{definition}
\label{def:coupledpotential}
The \emph{coupled-system potential} can be computed for general vector recursions written in the form of  \eqref{eq:spatial_fp}.
Integrating a scaled version of the vector update step along a curve $\mathcal{C}$, from $\bm{0}$ to $\x$, gives the potential function
\begin{align*}
  U(\x;\epsilon) &= \int_{\mathcal{C}}\g'(\z) (\z -
  \A ^{\intercal}\f(\A \g(\z); \epsilon))\cdot \mathrm{d}\z \\
  &=  \g(\x)^{\intercal}\x - G(\x) - F(\A \g(\x);\epsilon),
\end{align*}
where $\g'(\x) = \text{diag}([g'(x_i)])$, $G(\x) = \int_{\mathcal{C}}
\g(\z)\cdot\mathrm{d}\z = \sum_i G(x_i)$ and $F(\x; \epsilon) =
\int_{\mathcal{C}} \f(\z; \epsilon)\cdot\mathrm{d}\z = \sum_i F(x_i; \epsilon)$.
\end{definition}

\begin{remark}
A key observation in this paper is that a potential function for coupled systems can be written in the simple form in Def.~\ref{def:coupledpotential}.
Remarkably, this holds for general coupling coefficients because of the $\A,\A^{\intercal}$ reciprocity that appears naturally in SC.
\end{remark}

\begin{lemma}
\label{lem:shiftnormbound}
Let $\x \! \in \! [0,1]^n$ be a non-decreasing vector generated by averaging $\z\in[0,1]^n$ over a sliding window of size $w$.
Let the shift operator $\bm{S}:\mathbb{R}^{n}\to\mathbb{R}^{n}$ be defined by $\left[\bm{S} \x\right]_{1}=0$ and $\left[\bm{S} \x\right]_{i}=x_{i-1}$ for $i\geq2$.
Then, one can show $\norm{\bm{S}\x-\x}_{\infty} \leq \frac{1}{w}$ and $\norm{\bm{S}\x-\x}_1 = x_n = \| \x \|_\infty$.
\end{lemma}
\begin{IEEEproof}
The bound $\norm{\bm{S}\x-\x}_{\infty} \leq \frac{1}{w}$ follows from
\begin{align*}
x_{i}-x_{i-1} &= \tfrac{1}{w} \sum\nolimits_{j=0}^{w-1} z_{i-j} - \tfrac{1}{w} \sum\nolimits_{j=0}^{w-1} z_{i-1-j} \leq \tfrac{1}{w},
\end{align*}
where $z_i=0$ for all $i<0$. Since $\x$ is non-decreasing, the
1-norm sum telescopes and we get $ \| \bm{S} \x- \x \|_1 =
|0-x_{1}|+\sum_{i=2}^{n}|x_{i-1}-x_{i}|=x_{n} = \| \x \| _{\infty}$.
\end{IEEEproof}

\begin{lemma}
\label{lem:shiftenergy}
For the vector one-sided SC system, a shift changes the potential by $ U(\bm{S}\x;\epsilon) - U(\x;\epsilon)=-U(x_{i_0};\epsilon)$.
\end{lemma}
\begin{IEEEproof}[Sketch of Proof]
Rewriting the coupled potential gives
\[ U(\x;\epsilon) = \sum\nolimits_{i=-L-w}^{2w+i_0} \left[ g(x_i)x_i - G(x_i) - F\left( [\A\g(\x)]_i ; \epsilon \right) \right]. \]
One can verify that the contribution, from the first two terms in the square brackets, to $U(\bm{S}\x;\epsilon) - U(\x;\epsilon)$ is a telescoping sum that leaves only the difference between the first and last values.
For the third term in the square brackets, more care is required.
Since the first $w$ values of $\x$ are 0 and the last $2w+1$ values of $\x$ equal $x_{i_0}$,
it can be shown that $\sum_{i=-L-w}^{2w+i_0}
F\left([\A\g(\bm{S}\x)]_i; \epsilon \right) - F\left([\A\g(\x)]_i;
\epsilon \right) = F( g(0); \epsilon )- F\left(g(x_{i_0}); \epsilon
\right)$. Thus, we have
\[ \!\!\!\!\!U(\bm{S}\x;\epsilon) - U(\x;\epsilon) = -U(x_{i_0};\epsilon). \vspace{-6.5mm}\]
\end{IEEEproof}
\vspace{2mm}

\begin{lemma}
  For the SC potential, the norm of the Hessian $ U''(\x;\epsilon)$ is independent of $L$ and $w$ and
satisfies
\[\norm{U''(\x;\epsilon)}_{\infty} \leq K_{f,g} \triangleq \|g'\|_\infty + \|g'\|_\infty^2 \|f'\|_\infty  + \|g''\|_\infty, \]
where $\| h \|_\infty = \sup_{x\in[0,1]} \abs{h(x)}$ for functions $h:[0,1]\to\mathbb{R}$.
\end{lemma}
\begin{IEEEproof}
One can verify that the Hessian is given by
\begin{align*}
  U''(\x;\epsilon) &= \g'(\x) - (\A \g'(\x))^{\intercal}\f'(\A \g(\x);\epsilon)\A \g'(\x) \\
  &\phantom{=} + \g''(\x) \text{diag}\left(\x - \A ^{\intercal}\f(\A \g(\x);\epsilon)\right),
\end{align*}
where $\g''(\x) = \text{diag}([g''(x_i)])$.
Taking the norm gives
\begin{align*}
  &\norm{U''(\x;\epsilon)}_{\infty} \leq \norm{\g'(\x)}_{\infty} + \norm{\A \g'(\x))}_{1}\norm{\f'(\A \g(\x);\epsilon)}_{\infty}\\
&\phantom{==}\cdot\norm{\A \g'(\x))}_{\infty} +\norm{\g''(\x) \text{diag}\left(\x -
  \A ^{\intercal}\f(\A \g(\x);\epsilon)\right)}_{\infty}.
\end{align*}
Since $\norm{\A}_{\infty} \!=\! \norm{\A}_{1} \!=\! 1$ and $\norm{\g'(\x)}_{\infty} \! = \! \norm{\g'(\x)}_1 \leq \| g' \|_\infty$, we find that
$\norm{U''(\x;\epsilon)}_{\infty} \leq \|g'\|_\infty \!+ \|g'\|_\infty^2 \|f'\|_\infty \!+ \|g''\|_\infty$.
\end{IEEEproof}

We now state the main result of the paper.
Roughly speaking, it says that, if $\epsilon < \epsilon^*$ and $w$ is sufficiently large, then one can always lower the coupled potential of a non-zero vector by shifting.
Since this implies the next step of the recursion must reduce some value, the only fixed point is the zero vector.
\begin{theorem}
\label{thm:main_theorem}
 Consider a scalar admissible system~$(f,g)$.
 If $\epsilon < \epsilon^*$ and $w > K_{f,g}/\Delta E(\epsilon)$, then the only fixed point of the spatially-coupled system, defined by
 \eqref{eq:4}, is $\x=\bm{0}$.
\end{theorem}
\begin{IEEEproof}
  Using Lem.~\ref{lem:monotone}, let $\x$ be the unique fixed point of the one-sided recursion defined in Def.~\ref{def:OneSidedSC}.
  This fixed point upper bounds the fixed point of the
  basic SC system defined in Def.~\ref{def:SpatiallyCoupled}.  If $\x\neq 0$, then $x_{i_0}\geq u (\epsilon)$
  because the system has no fixed points with $x_i < u (\epsilon)$ for all $i$.
  From Lem.~\ref{lem:shiftenergy}, we have $\Delta U \triangleq U(\bm{S}\bm{x};\epsilon) - U(\bm{x};\epsilon)= -U(x_{i_0};\epsilon)$.
  Expanding $U(\bm{S}\x;\epsilon)$ in a Taylor series (with remainder) around $U(\x;\epsilon)$ gives
\begin{align*}
 U'(&\bm{x};\epsilon)\cdot(\bm{S}\bm{x}-\bm{x}) =  U(\bm{S}\bm{x};\epsilon) - U(\bm{x};\epsilon) \\
  &\phantom{=} - \int_0^1(1-t)(\bm{S}\bm{x}-\bm{x})^{\intercal}U''(\bm{x}(t);\epsilon)(\bm{S}\bm{x}-\bm{x})\mathrm{d}t \\
  &\leq \Delta U + \left| \int_0^1(1-t)(\bm{S}\bm{x}-\bm{x})^{\intercal}U''(\bm{x}(t);\epsilon)(\bm{S}\bm{x}-\bm{x})\mathrm{d}t \right| \\
  &\leq \Delta U + \norm{\bm{S}\bm{x}-\bm{x}}_1\max_{t\in[0,1]}\norm{U''(\bm{x}(t);\epsilon)}_{\infty}\norm{\bm{S}\bm{x}-\bm{x}}_{\infty} \\
  &\leq  -U(x_{i_0};\epsilon) + \tfrac{1}{w} x_{i_0}\max_{t\in[0,1]}\norm{U''(\bm{x}(t);\epsilon)}_{\infty} \\
  &\leq  -U(x_{i_0};\epsilon) + \tfrac{1}{w} K_{f,g} \\
  &<  -U(x_{i_0};\epsilon) + \Delta E (\epsilon) \leq 0,
\end{align*} \vspace*{-0.25mm}%
where the last steps hold because $w > K_{f,g}/\Delta E (\epsilon)$, $x_{i_0} \leq 1$, and $U(x;\epsilon) \geq \Delta E (\epsilon)$ for $x \geq u (\epsilon)$.

%

Now, we observe that $\bm{S}\x-\x \preceq \bm{0}$ (i.e., the fixed point is non-decreasing) and $[\bm{S}\x-\x]_i$ is zero for $i\notin \mathcal{L}'$.
So, $U'(\x;\epsilon)$ is positive in at least one component (i.e., there exists $i\in \mathcal{L}'$ such that $[U'(\x;\epsilon)]_i > 0$).
Since $g'(x)\geq 0$ and $[U'(\x;\epsilon)]_i = g'(x_i)[\x -
\A^{\intercal}f(\A g(\x);\epsilon)]_i$, it follows that $g'(x_i) > 0$
and $[\A^{\intercal}f(\A g(\x);\epsilon)]_i < x_i$.
So, one more iteration must reduce the value of the $i$-th component for some $i\in\mathcal{L}'$.
This gives a contradiction and shows that the only fixed point of the one-sided SC system is $\x=\bm{0}$.
The result follows since the fixed point of the basic SC system is
upper bounded by this f.p.
\end{IEEEproof}


\vspace{-0.60mm}
\section{Applications}
\label{sec:applications}

In this section, we consider some applications of
Theorem~\ref{thm:main_theorem} for coding problems that  are
characterized by a scalar recursion. We liberally use notation and
definitions from \cite{RU-2008}.

\vspace{-0.75mm}

\subsection{Irregular LDPC Codes}
\label{sec:regular-ldpc-codes}

\vspace{-0.25mm}

Consider the ensemble LDPC$(\lambda,\rho)$ and transmission over an
erasure channel with parameter $\epsilon$. Let $x^{(\ell)}$ be the
fraction of erasure messages sent from variable to check nodes during iteration $\ell$. The
DE equation can be written in the form of~\eqref{eq:1}, where
$f(x;\epsilon) = \epsilon\lambda(x)$ and $g(x) = 1 -
\rho(1-x)$. It is easy to verify that $f$ and $g$ describe a scalar admissible system
if $\lambda(0)=0$. The single system potential is given by
\vspace{-0,25mm}
\begin{align}
\label{eq:ldpc_potential}
  U(x;\epsilon) = \tfrac{1}{L'(1)}\left(-P(x) + (\epsilon(x) - \epsilon)L(1-\rho(1-x))\right),
\end{align}
where $P(x)$ is the trial entropy defined in
\cite[Def. 3.119]{RU-2008}, $\epsilon(x) = x/\lambda(1 - \rho(1-x))$ and $L(x) =
\int_0^x\lambda(y)\text{d}y/\int_0^1\lambda(y)\text{d}y$. In this
case, the potential $U(x;\epsilon)$ is the same as the pseudo-dual of the Bethe
variational entropy in \cite[Part 2, pp.~62-65]{Vontobel-acorn09}.
\begin{lemma}
\label{lem:bec_ldpc}
  Consider the potential threshold $\epsilon^*$ given by
  (\ref{eq:3}). Let $\epsilon^{\text{Max}}$ be the Maxwell threshold
  \cite[Conj.~1]{Measson-it08}, defined by \vspace{-1mm}
  \begin{equation}
    \label{eq:6}
    \epsilon^{\text{Max}} = \min \left\{\epsilon(x) \, | \, P(x)=0,x\in[0,1]\right\}. \vspace{-1mm}
  \end{equation}
  Then, $\epsilon^* = \epsilon^{\text{Max}}$ for the ensemble
  LDPC$(\lambda,\rho)$.
\end{lemma}
\begin{IEEEproof}[Sketch of Proof]
  Let $x^{\mathrm{Max}}$ be the $x$-value that achieves the minimum.
  Then, \eqref{eq:ldpc_potential} and \eqref{eq:6} imply
  $U(x^{\mathrm{Max}};\epsilon^{\mathrm{Max}}) =
  U(x^{\mathrm{Max}};\epsilon(x^{\mathrm{Max}})) = -
  P(x^{\mathrm{Max}})/L'(1) = 0$.
  One can show $\Delta E(\epsilon^{\text{Max}}) =
  U(x^{\text{Max}};\epsilon^{\text{Max}}) = 0$, which implies 
  $\epsilon^*= \epsilon^{\text{Max}}$.
\end{IEEEproof}

\begin{remark}
For regular ensembles, $\epsilon^{\text{Max}}$ equals the MAP threshold $\epsilon^{\text{MAP}}$ and this is conjectured to hold in general.
\end{remark}

Consider an SC ensemble of irregular LDPC($\Lambda,P$) codes defined as follows. The $f$-nodes at
each position are replaced by $M$ copies of the node
degree profile $\Lambda(x) = \sum_i \Lambda_i x^i$, where $\Lambda_i$
is the number of bit nodes of degree $i$.  The $g$-nodes at each
position are replaced by $M$ copies of the node degree
profile $P(x) = \sum_i P_i x^i$, where $P_i$ is the number of check
nodes of degree $i$. For sufficiently large $M$, these nodes can be
coupled uniformly using an averaging window
of length $w$ (see Fig.~\ref{fig:sc_system}) in a manner similar to
the $(\dl,\dr,L,w)$ ensemble defined in~\cite{Kudekar-it11}.

\begin{corollary}
Applying Theorem~\ref{thm:main_theorem} shows that, if $\epsilon < \epsilon^{\text{Max}}$ and $w >
  K_{f,g}/\Delta E(\epsilon)$, then
the SC DE recursion converges to the zero vector.
\end{corollary}


\vspace{-1mm}

\subsection{Generalized LDPC Codes on the BEC and BSC}
\label{sec:gldpc-codes}

Consider a generalized LDPC (GLDPC) code with degree-2 bits and generalized check constraints based on a BCH code of block-length $n$.
For iterative decoding using bounded-distance decoding of the BCH code, the DE recursions can be derived for both the BEC and binary symmetric channel (BSC)~\cite{Jian-isit12}.
On the BEC, the code is chosen to correct all patterns of at most $t$ erasures and, on the BSC, the code is chosen to correct all error patterns of weight at most $t$.

For both cases, the iterative decoding performance of this ensemble is characterized by a DE recursion
of the form~(\ref{eq:1}), where $\epsilon$ denotes the channel parameter.
In this case, the scalar system is defined by $f(x;\epsilon)\triangleq \epsilon x$ and $g(x)\triangleq \sum_{e=t}^{n-1}\binom{n-1}{e}x^e(1-x)^{n-1-e}$.
Here, $x$ denotes the erasure (resp. error) probability of bit-to-check messages for the BEC (resp. BSC) case.

It can be verified that $f$ and $g$ define a scalar admissible system whose
single-system potential is given by $U(x;\epsilon) = \tfrac{1}{2}\left(-P(x) + g(x)(x- f(g(x);\epsilon))\right)$, where \vspace{-0.5mm}
\[ P(x) = \textstyle{\int_0^x} (g(z))^2 \epsilon' (z) \mathrm{d}z = -xg(x) + 2 \textstyle{\int_0^x} g(z) \mathrm{d}z \]
is the analogous to the trial entropy defined in~\cite{RU-2008}.

\begin{lemma}
For $2 \leq t \leq \lfloor\frac{n-1}{2}\rfloor$ on the BEC, $P(x)$
has an unique root, $\bar{x}$, in $(0,1]$. Let $\bar{\epsilon}
\triangleq \epsilon(\bar{x})$ be the $\epsilon$-root, and
$\epsilon^*$ be the potential threshold defined by (\ref{eq:3}).
Then, $\epsilon^*=\bar{\epsilon}$ for the GLDPC ensemble.
\end{lemma}

\begin{remark}
Since this decoder uses suboptimal component decoders, the threshold defined by the unique zero of $P(x)$ does not give an upper bound on the MAP threshold.
\end{remark}

We now describe the SC ensemble for the ensemble of GLDPC codes. The
$f$-nodes in Fig.~\ref{fig:sc_system} are replaced by $Mn$
degree-$2$ variable nodes and the $g$-nodes are replaced by $2M$ BCH
codes of block-length $n$. These nodes are coupled using an
averaging window of length $w$ in a manner similar to the
$(\dl,\dr,L,w)$ ensemble defined in~\cite{Kudekar-it11} (e.g.,
see~\cite{Jian-isit12}).
\begin{corollary}
Applying Theorem~\ref{thm:main_theorem} to this SC GLDPC ensemble
shows that the SC DE recursion converges to the zero vector whenever
$\epsilon<\bar{\epsilon}$ and $w > K_{f,g}/ \Delta E(\epsilon) $.
\end{corollary}


\vspace{-0.5mm}

\subsection{Intersymbol-Interference Channels with Erasure Noise}
\vspace{-0.5mm}
\label{sec:gec-ldpc}
In~\cite{Pfister-jsac08}, a family of intersymbol-interference (ISI)
channels with erasure noise is investigated as an analytically
tractable model of joint iterative decoding of LDPC codes and channels
with memory.  Let $\psi(t;\epsilon)$ be the function that maps the
\emph{a priori} erasure rate $t$ from the code and the channel erasure
rate $\epsilon$ to the erasure rate of extrinsic messages from the
channel detector to the bit nodes. Then, the
resulting DE update equation for the erasure rate, $x^{(\ell)}$, of
bit-to-check messages can be written in the form of (\ref{eq:1}),
where $f(x;\epsilon) = \psi(L(x);\epsilon)\lambda(x)$ and $g(x) = 1 -
\rho(1-x)$~\cite{Pfister-jsac08}.
Under mild conditions on $\psi(x;\epsilon)$,
this defines a scalar admissible system with potential
\vspace{-0.75mm}
\begin{align*}
  U(x;\epsilon) = \tfrac{1}{L'(1)} \Psi(L(g(x));\epsilon(x)) - \Psi(L(g(x));\epsilon) - P(x), \vspace*{-1.75mm}
\end{align*}
where $\epsilon(x)$ and $P(x)$ are defined in \cite{Nguyen-arxiv11} for
generalized erasure channels and $\Psi(x;\epsilon) =
\int_0^x\psi(y;\epsilon)\text{d}y$.
\begin{lemma}
  Consider the potential threshold $\epsilon^*$ defined by
  (\ref{eq:3}) and let $\epsilon^{\text{Max}}$ be defined by
  (\ref{eq:6}). Then, $\epsilon^* = \epsilon^{\text{Max}}$.
  If, in addition, $P(x)$ has a unique root $\bar{x} \in (0,1]$, then
  $\bar{\epsilon}^{\text{MAP}}=\epsilon(\bar{x})$ is an upper bound on the
  MAP threshold and $\epsilon^* = \bar{\epsilon}^{\text{MAP}}$.
\end{lemma}
\begin{IEEEproof}
Omitted due to similarity with Lem.~\ref{lem:bec_ldpc}.
\end{IEEEproof}
\begin{corollary}
  Consider the SC ensemble defined in
  Section~\ref{sec:regular-ldpc-codes}. If $\epsilon <
  \epsilon^*$ and $w >
  K_{f,g}/\Delta E(\epsilon)$, then
  Theorem~\ref{thm:main_theorem} shows that the SC DE recursion
  converges to the zero vector.
\end{corollary}

\section{Conclusions}

A new theorem is presented that provides a simple proof of threshold saturation for many scalar DE recursions.
The conditions of the theorem are verified for the density-evolution (DE) equations of irregular LDPC codes on the BEC, a class of generalized LDPC codes, and the joint iterative decoding of some intersymbol-interference channels with erasure noise.
Therefore, threshold saturation is now proved for these cases.
Moreover, we believe this approach opens the door to threshold-saturation proofs for many more general spatially-coupled systems.
\vspace{-0.15in}

\bibliographystyle{ieeetr}
\bibliography{WCLabrv,WCLbib,WCLnewbib}

\end{document}